\documentclass[article,a4paper]{raa}           
\usepackage{graphicx,times}             
\usepackage{natbib}
\usepackage{amssymb,amsmath}

\bibpunct{(}{)}{;}{a}{}{,}

\usepackage[pagebackref=true]{hyperref}

\begin{document}

  \title{High-resolution Solar Image Reconstruction Based on Non-rigid Alignment}

   \volnopage{Vol.0 (20xx) No.0, 000--000}      
   \setcounter{page}{1}          

   \author{Hui Liu 
   \and Zhenyu Jin
   \and Yongyuan Xiang
   \and Kaifan Ji
   }

   \institute{Yunnan Observatories, Chinese Academy of Sciences, Kunming, Yunnan 650216, People's Republic of China.; {\it liuhui@ynao.ac.cn}\\
\vs\no
   {\small Received 20xx month day; accepted 20xx month day}}

\abstract{ Suppressing the interference of atmospheric turbulence and obtaining observation data with a high spatial resolution is an issue to be solved urgently for ground observations. One way to solve this problem is to perform a statistical reconstruction of short-exposure speckle images. Combining the rapidity of \textit{Shift-Add} and the accuracy of \textit{speckle masking}, this paper proposes a novel reconstruction algorithm-NASIR (Non-rigid Alignment based Solar Image Reconstruction). NASIR reconstructs the phase of the object image at each frequency by building a computational model between geometric distortion and intensity distribution and reconstructs the modulus of the object image on the aligned speckle images by \textit{speckle interferometry}.
We analyzed the performance of NASIR by using the correlation coefficient, power spectrum, and coefficient of variation of intensity profile (\emph{CVoIP}) in processing data obtained by the NVST (1m New Vacuum Solar Telescope).
The reconstruction experiments and analysis results show that the quality of images reconstructed by NASIR is close to \textit{speckle masking} when the seeing is good, while NASIR has excellent robustness when the seeing condition becomes worse.
Furthermore, NASIR reconstructs the entire field of view in parallel in one go, without phase recursion and block-by-block reconstruction, so its computation time is less than half that of \textit{speckle masking}. Therefore, we consider NASIR is a robust and high-quality fast reconstruction method that can serve as an effective tool for data filtering and quick look.
}

   \authorrunning{H. Liu et al.}            
   \titlerunning{High-resolution Solar Image Reconstruction Based on Non-rigid Alignment}  

   \maketitle

%
%
\section{Introduction}           
\label{sect:intro}

At present, solar physics research has entered a period of small scale and fine structure. High-resolution observational images are urgently needed. However, diffraction-limited imaging cannot be achieved with ground-based optical telescopes due to the interference of the turbulent atmosphere.
\par To mitigate the adverse effects of the atmosphere and achieve near-diffraction-limited imaging, the researchers propose the speckle imaging method for high-resolution reconstruction by statistical analysis. The speckle imaging method started from the pioneering work of Labeyrie \citep{1970Attainment}, and its processing object is the speckle images. Speckle images are short-exposure images taken with a telescope with an aperture much larger than the atmospheric coherence length. The speckle imaging method can obtain high-resolution reconstructed images close to the diffraction limit of the telescope and has a relatively complete theoretical support.
\par Speckle imaging methods can roughly be classified into spatial domain and frequency domain reconstruction methods. Spatial reconstruction methods directly reconstruct the spatial structure and intensity of the object. The characteristics of the spatial domain method are simple algorithms and clear spatial descriptions. The typical representative of the spatial domain reconstruction method is \textit{Shift-Add} \citealt{BATES1980365}.
Frequency domain reconstruction methods usually reconstruct the phase and modulus separately. Phase is reconstructed by \textit{speckle masking} \citep{lohmann1983speckle}, while modulus is by \textit{speckle interferometry} \citep{1970Attainment}.
Most of the above-mentioned classic algorithms originated in the 1970s and 1980s, and subsequent researchers also have substantial and valuable research works to improve these methods \citep{jin2008high,XIANG20168,2010PABei..28...72H,1999AcOpS..19..935L}. However, there are still issues that need to be further studied and resolved in practical applications.The main issue of spatial domain methods is to align the regions containing object diffraction imaging information accurately. Taking \textit{Shift-Add} as an example, the algorithm assumes that the maximum intensity points of the speckle images coincide with that of the object image and performs rigid translation (uniform displacement of the entire image) and superimposition of each speckle image to obtain a reconstructed image.
However, for an extended source like \textit{Sun}, due to the influence of the surrounding structure, the distribution of maximum intensity points will change significantly under the influence of the atmosphere.
Not only exist multiple points of maximum intensity, but their positions will also deviate. The assumption that points of maximum intensity coincide each other no longer holds. In addition, the overall displacement of the image ignores the distortion differences caused by different fields of view and frequency components, which makes it difficult to obtain satisfactory reconstruction results with the traditional rigid \textit{Shift-Add}.
\par Speckle imaging frequency-domain reconstruction algorithm is currently the most widely used high-resolution reconstruction method \citep{1999AcOpS..19..935L}. Well-known ground-based solar telescopes such as NVST \citep{liu2014new} and GST (Goode Solar Telescope) \citep{cao2010scientific} use this method to reconstruct observational data. A typical representative of the frequency domain reconstruction method is \textit{speckle masking}.
\par Since \textit{speckle masking} estimates the phase of the object based on the triple correlation of the speckle images, that is, recursively from the low-frequency phase to the high-frequency, it is inevitably affected by noise such as photon noise during the recursion process. Due to the recursion process, the noise influence will accumulate in the high-frequency part, and in severe cases, it will distort the fine structure of the object. Therefore, the performance of \textit{speckle masking} will significantly reduce with the decrease in seeing \citep{1999AcOpS..19..935L}.
\par Therefore, we propose a new high-resolution solar image reconstruction algorithm named NASIR in this paper. NASIR draws on the basic theoretical models of \textit{speckle masking} and \textit{Shift-Add} and combines the ideas of the \textit{optical flow} method \citep{horn1981determining} in computer vision. The algorithm obtains the distortion displacement field of each frame of the speckle image relative to the reference image (RI) by establishing a computational model between the intensity change and the displacement field. And the distortion is corrected by pixel-by-pixel non-rigid alignment to obtain the phase estimation of the object image. The modulus of the object image is by \textit{speckle interferometry} on the aligned speckle images.
\par Section 2 of the paper introduces the principle and realization of NASIR. Section 3 applies the algorithm to reconstruct the observation data of NVST and compares it with the classical methods. Section 4
discusses the features of NASIR and issues that need to be further studied. Our conclusions are in Section 5.


\section{The Principle and Implementation of NASIR}
     \label{S-The Principle and Implementation of NASIR}
Similar to \textit{speckle masking}, NASIR also reconstructs phase and modulus separately. For phase reconstruction, NASIR obtains the distorted displacement field of each frame of the speckle image relative to the reference image (RI) by establishing a computational model between the intensity variation and the displacement field. The phase estimation of the object image is by correcting the distortion by pixel-by-pixel non-rigid alignment. The modulus is by speckle interferometry on the aligned speckle images.
\par The above process involves displacement field estimation, band-pass filtering, iterative alignment, and modulus reconstruction of the object image.

\subsection{Displacement Field Estimation Based on Intensity Changes} 
  \label{S-Displacement Field Estimation Based on Intensity Changes}
Based on the idea of the \textit{optical flow}, we established the pixel displacement and intensity variation model of the speckle image. With this model, the phase distortion caused by atmospheric turbulence is corrected by reverse displacement correction.
For the convenience of analysis and calculation, we first use a polynomial to approximate the relationship between pixel intensity and pixel coordinates in a local region of the image. Taking the quadratic polynomial as an example, the intensity distribution of the local area in a speckle image can express approximately as:
\begin{equation}\label{1}
f(\boldsymbol{x})\approx \boldsymbol{x}^T\boldsymbol{A}\boldsymbol{x}+\boldsymbol{b}^T\boldsymbol{x}+c
\end{equation}

Where $\boldsymbol{x}$ is the coordinates of a local region in a speckle image, $f(\boldsymbol{x})$ is the pixel intensity at $\boldsymbol{x}$, $\boldsymbol{A}$ is a symmetric matrix, $\boldsymbol{b}$ is a displacement vector, and c is a scalar. These parameters can be estimated by the \textit{weighted least square} method.

Assume that the approximate expression formula of the corresponding region of the reference image \textbf{RI} (This article uses the ensemble average of the speckle images as the initial reference image) is:
\begin{equation}\label{2}
f_1(\boldsymbol{x})\approx\boldsymbol{x}^T\boldsymbol{A_1}\boldsymbol{x}+\boldsymbol{b_1}^T\boldsymbol{x}+c_1
\end{equation}
If the speckle image has an ideal translation $\boldsymbol{d}$ with respect to \textbf{RI} in this region, the relationship between the intensity of this area and the spatial coordinate of the speckle image can be expressed as:

\begin{eqnarray}\label{3}
f_2(\boldsymbol{x})~&=& f_1(\boldsymbol{x}-\boldsymbol{d})  \nonumber    \\
~&=& (\boldsymbol{x}-\boldsymbol{d})^T\boldsymbol{A_1}(\boldsymbol{x}-\boldsymbol{d})+\boldsymbol{b_1}^T(\boldsymbol{x}-\boldsymbol{d})+c_1  \nonumber    \\
~&=& \boldsymbol{x}\boldsymbol{A_1}\boldsymbol{x}+(\boldsymbol{b_1}-2\boldsymbol{A_1}\boldsymbol{d})^T\boldsymbol{x}+\boldsymbol{d}^T\boldsymbol{A_1}\boldsymbol{d}-\boldsymbol{b_1}^T\boldsymbol{d}+c_1  \nonumber    \\
~&=&\boldsymbol{x}^T\boldsymbol{A_2}\boldsymbol{x}+\boldsymbol{b_2}^T\boldsymbol{x}+c_2
\end{eqnarray}

By comparing the coefficients of $f_1(\boldsymbol{x})$ and $f_2(\boldsymbol{x})$, we can get:

\begin{eqnarray}\label{4-6}
\boldsymbol{A_2}~&=& \boldsymbol{A_1}  \\
\boldsymbol{b_2}~&=& \boldsymbol{b_1} - 2\boldsymbol{A_1}\boldsymbol{d}  \\
c_2~&=& \boldsymbol{d}^T\boldsymbol{A_1}\boldsymbol{d}-\boldsymbol{b_1}^T\boldsymbol{d}+c_1
\end{eqnarray}

In the non-singular case of $\boldsymbol{A_1}$, we can obtain the translation $\boldsymbol{d}$ through equation (5):

\begin{equation}\label{7}
2\boldsymbol{A_1}\boldsymbol{d}=-(\boldsymbol{b_2}-\boldsymbol{b_1})
\end{equation}

\begin{equation}\label{8}
\boldsymbol{d}=(1/2)\boldsymbol{A_1^{-1}}(\boldsymbol{b_2}-\boldsymbol{b_1})
\end{equation}

So we established a computational model between pixel intensity and displacement under ideal translation. To more accurately describe the projection of the distortion on the image, a reasonable model of the displacement field can be obtained through optimization strategies such as the affine transformation of the displacement field and the weighting of neighborhood pixels  \citep{farneback2003two}. The specific method is as follows: for the projection of the general three-dimensional motion on the image plane, we use the affine transformation model to express the displacement field as:
\begin{equation}\label{9}
\begin{array}{l}
\boldsymbol{d_x}(x,y)=a_1+a_2x+a_3y+a_7x^2+a_8xy \\
\boldsymbol{d_y}(x,y)=a_4+a_5x+a_6y+a_7xy+a_8y^2
\end{array}
\end{equation}
Written in matrix form:
\begin{equation}\label{10}
\boldsymbol{d}=\boldsymbol{S}\boldsymbol{P}
\end{equation}

among them,
\begin{equation}\label{11}
\boldsymbol{S}=
 \left(
 \begin{array}{cccccccc}
     1 & x & y & 0 & 0 & 0 & x^2 & xy \\
     0 & 0 & 0 & 1 & x & y & xy  & y^2
 \end{array}
 \right)
 \end{equation}

\begin{equation}\label{12}
\boldsymbol{P}=
\left(
\begin{array}{cccccccc}
 a_1 & a_2 & a_3 & a_4 & a_5 & a_6 & a_7 & a_8
\end{array}
\right)
\end{equation}

then:
\begin{equation}\label{13}
\boldsymbol{P}=(\sum\limits_{i}{w_i\boldsymbol{S_i^T}\boldsymbol{A_i^T}\boldsymbol{A_i}\boldsymbol{S_i}})^{-1}\sum\limits_{i}({w_i\boldsymbol{S_i^T}\boldsymbol{A_i^T}\Delta\boldsymbol{b_i}})
\end{equation}

Where $i$  is the pixels in the neighborhood, $w_i$ is the weights of the pixel $i$. Neighborhood size needs to determine according to seeing ($\boldsymbol{r_0}$) (which will discuss in Section 4.2).
Solving $\boldsymbol{P}$ in equation (13) and substituting it into equation (10), we can obtain the displacement vector $\boldsymbol{d}$ of each pixel relative to \textbf{RI}.
Each speckle image can align to \textbf{RI} by inversely shifting and interpolating each pixel according to the $\boldsymbol{d}$ calculated at each time.

\subsection{Band-pass Filtering} 
  \label{S-Band-pass Filtering}
Due to the properties of atmospheric turbulence, the original speckle image is dominated by low-frequency energy. The displacement field calculated directly from the original speckle image mainly reflects the distortion of the low-frequency structure, which is not conducive to the recovery of the mid and high-frequency phase.
Therefore, based on the theoretical speckle interference transfer function (\textit{sitf}) \citep{korff1973analysis} and the optical transfer function (\textit{Otf}) of NVST \citep{XIANG20168}, we constructed a band-pass filter $\boldsymbol{H_{bp}}(\boldsymbol{u})$ as shown in Eq.14 to filter the original speckle image. This filter enhances mid and high-frequency features and suppresses noise outside the diffraction limit, thereby improving the quality of image alignment.


\begin{equation}\label{14}
\boldsymbol{H_{bp}}(\boldsymbol{u})=\frac{\textit{Otf}(\boldsymbol{u})}{\textit{sitf}(\boldsymbol{u})+\boldsymbol{S_n}}
\end{equation}

Where $\boldsymbol{S_n}$ is the noise power spectrum estimated from the speckle image sequence, and $\boldsymbol{u}$ is the spatial frequency.

\subsection{Reference Image Update and Iterative Alignment} 
  \label{S-Reference Image Update And Iterative Alignment}
Estimating the distortion displacement requires an image with a higher signal-to-noise ratio as the reference image \textbf{RI} (described in 2.1). The initial \textbf{RI} is the ensemble-average image of the speckle images filtered by Eq.14. After the speckle images align to the current \textbf{RI}, the signal-to-noise ratio of their ensemble-averaged image is improved, at this time, NASIR updates the \textbf{RI} with the newer ensemble-averaged image and performs displacement field estimation and alignment again.
\par After multiple iterations of alignment and updating, the quality of \textbf{RI} and the alignment accuracy of speckle images continuously improved. Then the phase of the ensemble-averaged image of the speckle images is used as the phase estimation of the object image.

\subsection{Modulus Reconstruction Based on Speckle Interferometry} 
  \label{S-Modulus Reconstruction Based on Speckle Interferometry}
The basic idea of \textit{speckle interferometry} is that when the exposure time is less than a certain threshold, the speckle images contain more information than the long-exposure image. The power spectrum of multiple consecutive short exposure speckle images can be time-averaged to restore the power spectrum of the object image \citep{1970Attainment}.
\par In practical observation, speckle images often contain various random noises. Therefore, NASIR uses the Wiener filter method of Eq.15 to reconstruct the modulus of the object image:

\begin{equation}\label{15}
|\boldsymbol{O}(\boldsymbol{u})|=\sqrt{\frac{<|\boldsymbol{I_i}(\boldsymbol{u})|^2>}{\textit{sift}(\boldsymbol{u})+\boldsymbol{S_n}}}\cdot{\textit{Otf}(\boldsymbol{u})}
\end{equation}

Where $|\boldsymbol{O}(\boldsymbol{u})|$ is the modulus of the object image, $<|\boldsymbol{I_i}(\boldsymbol{u})|^2>$ is the ensemble-average of the power spectrum of the speckle images, $\boldsymbol{S_n}$ is the noise power spectrum estimated from the speckle image sequence, $\textit{sift}(\boldsymbol{u})$ is calculated based on the seeing \citep{korff1973analysis}.
\par Since the distortions in speckle images are corrected, their ensemble-average images contain more mid-and high-frequency energy, which can further improve the contrast of the reconstructed image. So, unlike speckle masking, NASIR uses the aligned speckle image instead of the original speckle image to calculate $<|\boldsymbol{I_i}(\boldsymbol{u})|^2>$ and then obtains $|\boldsymbol{O}(\boldsymbol{u})|$ from Eq.(15).

\section{Reconstruction Results and Analysis}
     \label{S-Reconstruction Results and Analysis}
We compare the reconstruction performance of NASIR with that of correlation \textit{Shift-Add} and \textit{speckle masking} by reconstruction experiments on NVST observations.
\par NVST is the largest ground-based vacuum solar telescope in China, and it was put into use in 2010. The NVST has a clear aperture of 985 mm and is mainly used for high-resolution imaging observations of the solar photosphere, chromosphere, and spectrum \citep{liu2014new}. The main instruments of NVST include high-resolution multi-channel imaging systems \citep{xu2013primary} and high-dispersion multi-wavelength spectrometers\citep{wang2013first}.
\par NVST data products are divided into three levels: Level 0 is the original observation data; Level 1 is based on lucky imaging, corrected by dark current and flat field, and reconstructed by the \textit{Shift-Add}; Level 1+ is reconstructed by \textit{speckle masking }, which can produce reconstruction results close to the diffraction limit\citep{yan2020research}.
\par The experimental data set in this paper is derived from the Level 0 data of the NVST photosphere (TiO, center wavelength 705.8nm, bandwidth 1nm) and chromosphere (H$\alpha$, center wavelength 656.28$\pm$0.4nm, bandwidth 0.025nm) under different seeing degrees from 2013 to 2019. Each set of experimental data contains 100 short-exposure speckle images (Table 1). The seeing parameter $\boldsymbol{r_0}$ is calculated from the spectral ratio method\citep{von1984estimating}.
\begin{table}
\begin{center}
\caption{ Speckle images high-resolution reconstruction test data set:
}
\label{T1-simple}
\begin{tabular}{ccclr}     
  \hline                   
Data set No. & Band & Observation time & Active area & $\boldsymbol{r_0}$(cm) \\
  \hline
1 & TiO       & 2013-08-01 UT 03:44:53 & 11801  & 10.29 \\
2 & H$\alpha$ & 2019-10-04 UT 02:59:05 & 12749  & 10.78 \\
3 & TiO       & 2015-08-21 UT 10:01:57 & 12403  & 5.88 \\
4 & TiO       & 2019-04-11 UT 08:43:27 & 12738  & 4.09 \\
  \hline
\end{tabular}
\end{center}
\end{table}

\subsection{Analysis of Reconstruction Quality When Seeing is Good} 
  \label{S-Analysis of Reconstruction Quality When Seeing is Good}
Fig.1 Shows the reconstruction results of NVST TiO observation data (data set No.1 in Table 1) using \textit{correlation Shift-Add}, \textit{speckle masking}, and NASIR, respectively.
\begin{figure}[htbp]
  \begin{minipage}[t]{0.495\linewidth}
  \centering
   \includegraphics[width=60mm]{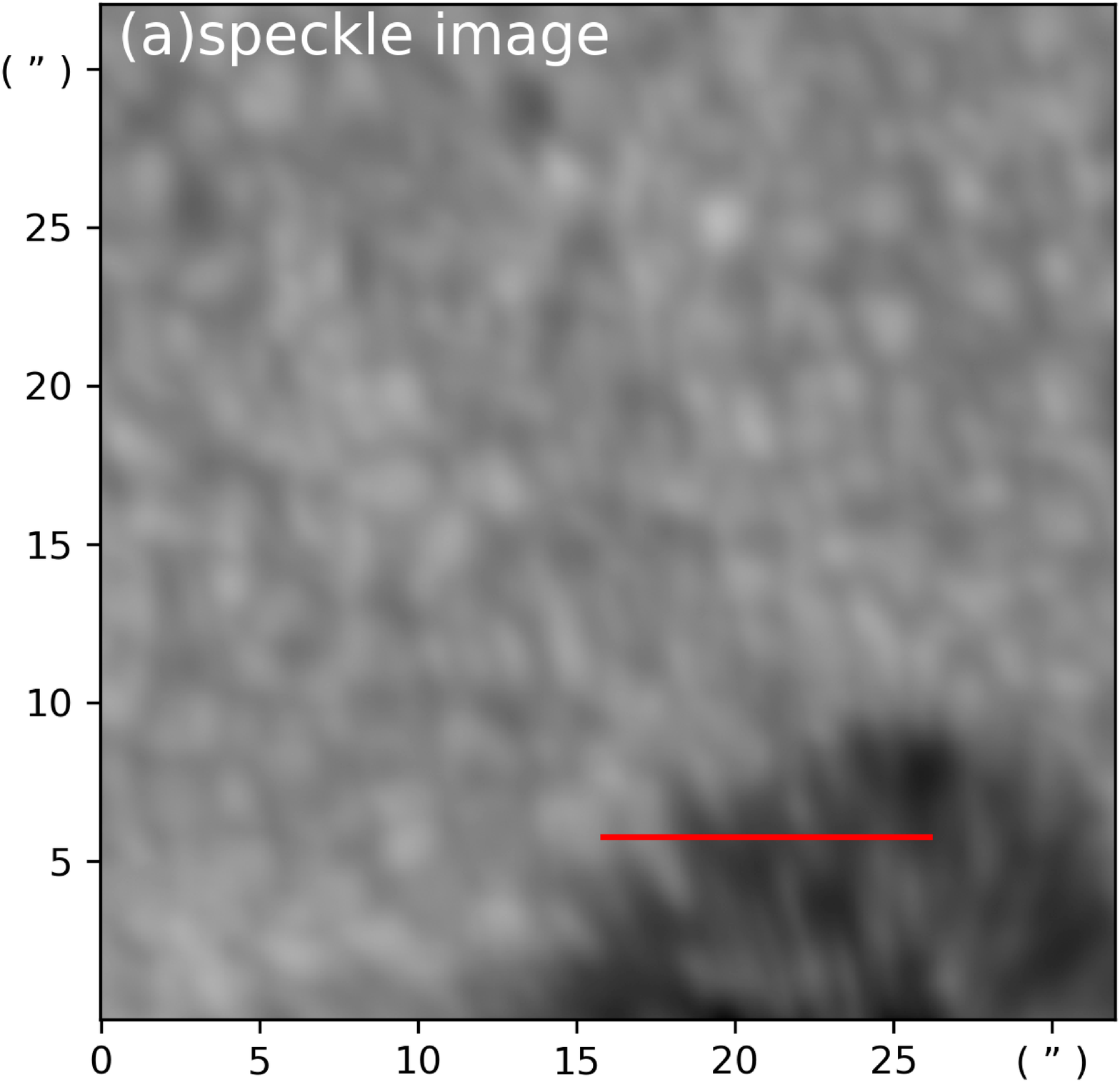}
  \end{minipage}%
  \begin{minipage}[t]{0.495\textwidth}
  \centering
   \includegraphics[width=60mm]{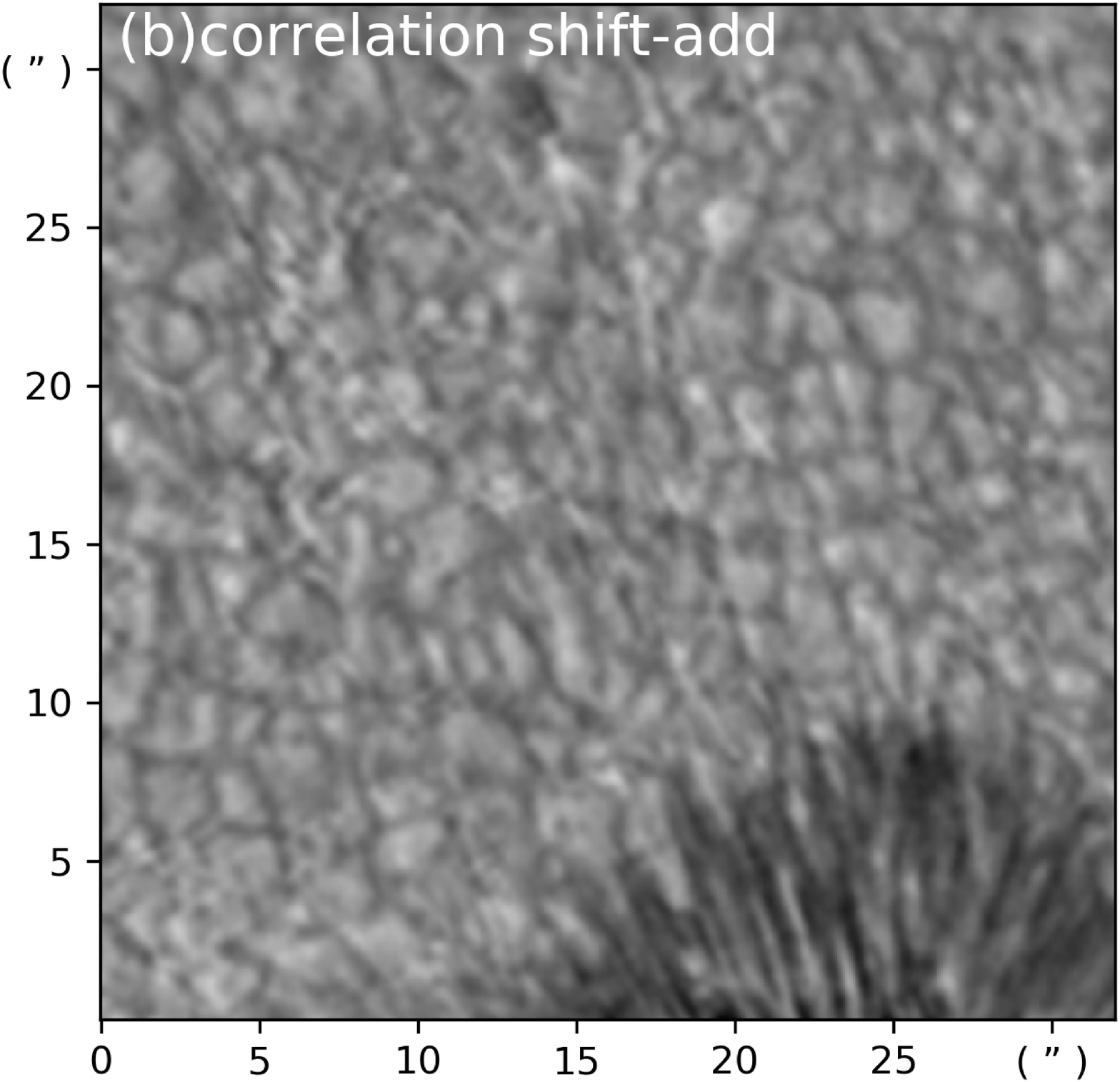}
  \end{minipage}%

  \begin{minipage}[t]{0.495\textwidth}
  \centering
     \includegraphics[width=60mm]{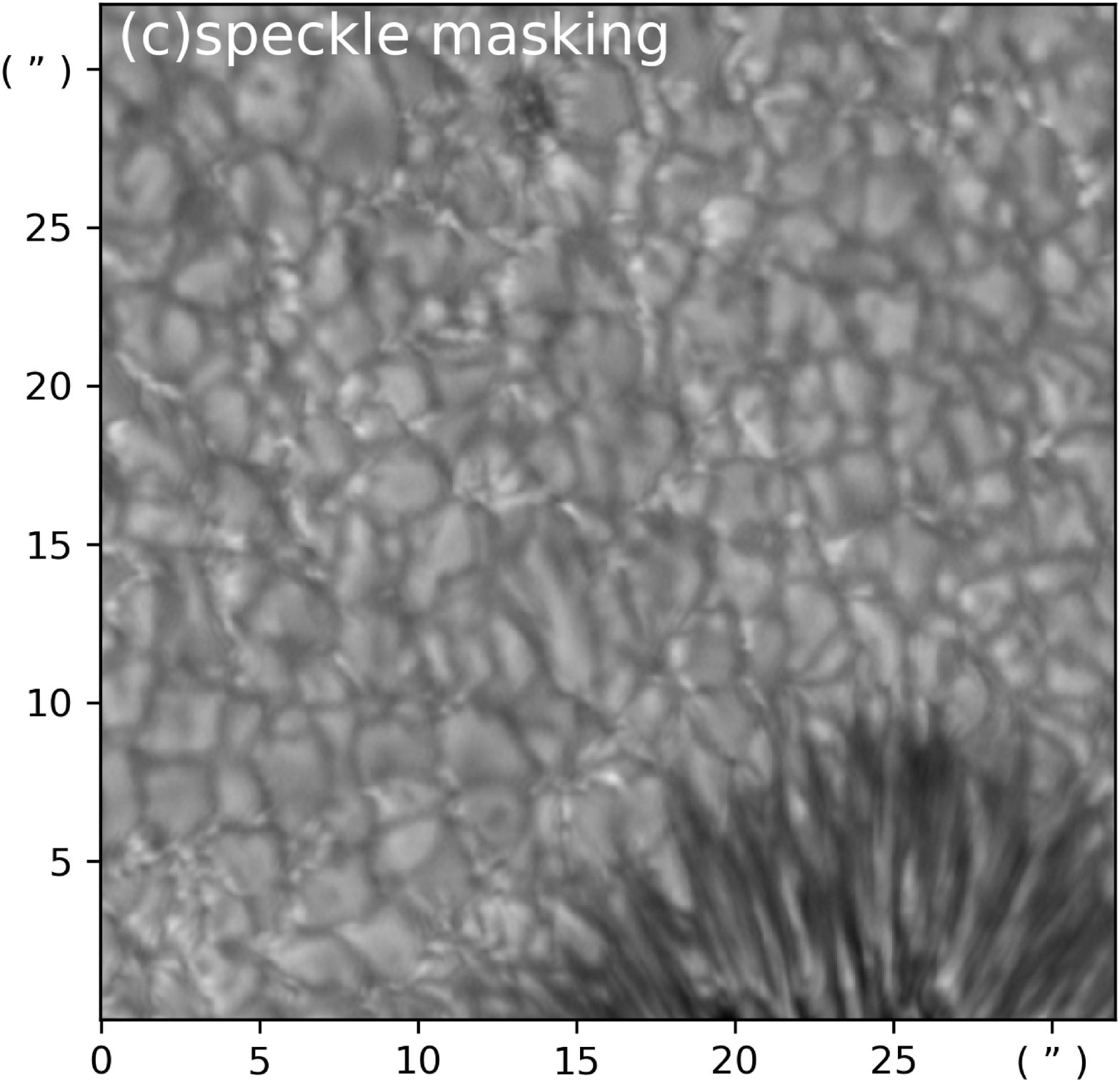}
  \end{minipage}%
  \begin{minipage}[t]{0.495\textwidth}
  \centering
   \includegraphics[width=60mm]{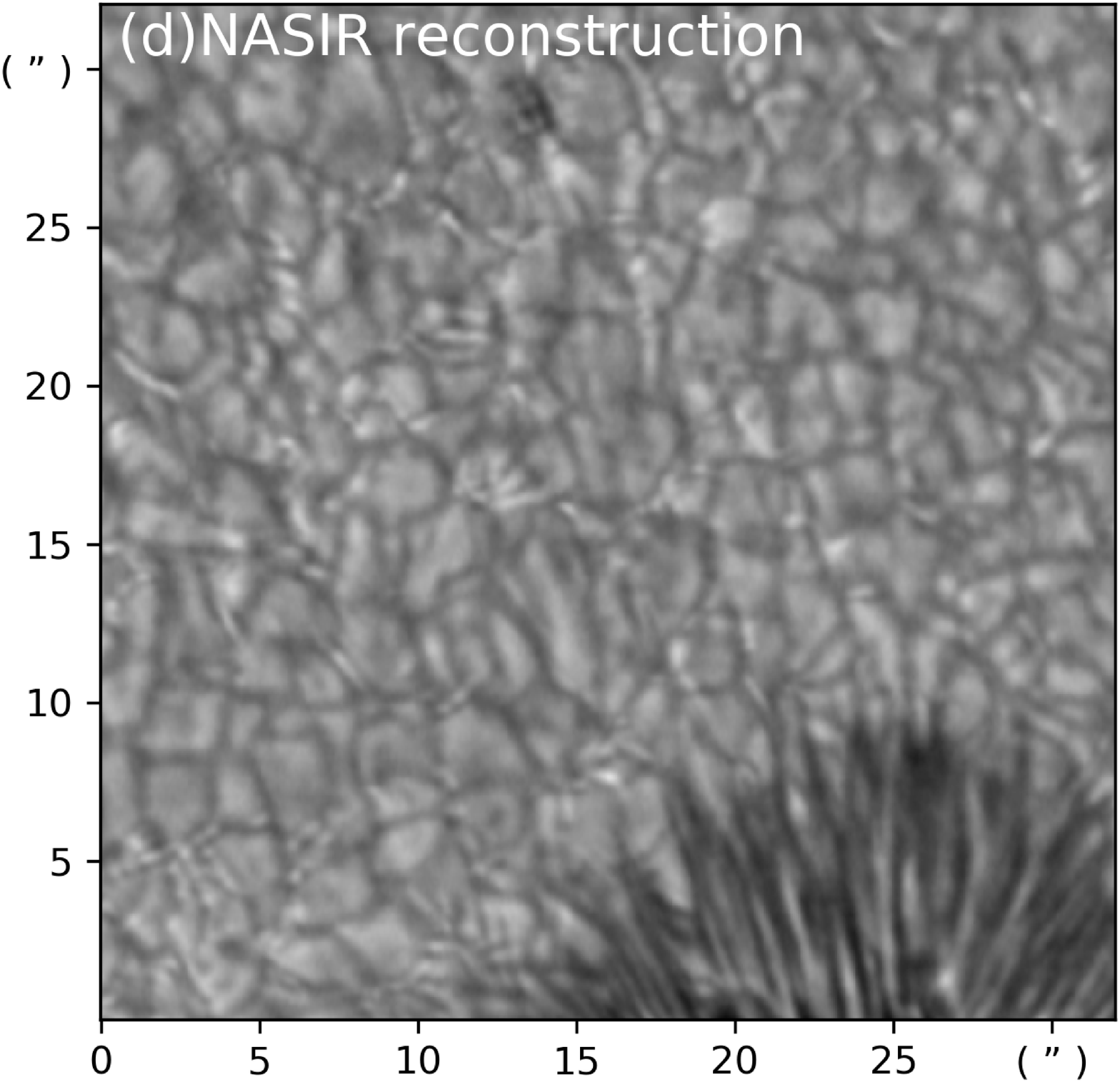}
  \end{minipage}%
  	  \caption{\label{Fig1} Reconstruction results of the NVST TiO observations with the active area 11801, 2013-08-01T:03:44:53. (a) A frame in a sequence of speckle images. (b) Reconstruction by \textit{correlation Shift-Add}. (c) Reconstruction by \textit{speckle masking}. (d) NASIR reconstruction.}
\end{figure}

It can be seen from Fig.1 (granule outlines, bright point and penumbral filaments of sunspot, etc.) that the reconstruction quality of NASIR is significantly better than that of \textit{Shift-Add}, and it is closer to the reconstruction quality of \textit{speckle masking}.

\par To intuitively compare the reconstruction quality in the frequency domain, we convert the 2-D power spectrum image to polar coordinates. Then, the power spectrum image in polar coordinates is averaged along the polar angular to obtain a 1-D power spectrum curve with the length of the polar diameter as the frequency and the average intensity as the energy.
\par The power spectrum curve in Fig. 2 (left) shows that the resolution of the NASIR reconstructed image is close to that of \textit{speckle masking} reconstructed image, about 0.15 arc-second, which is significantly higher than that of \textit{Shift-Add} reconstructed image of about 0.25 arc-second. The intensity profile in Fig.2 (right) further shows the similarity in intensity distribution between NASIR and \textit{speckle masking} reconstruction.

\begin{figure}[ht]
  \begin{minipage}[t]{0.495\linewidth}
  \centering
   \includegraphics[width=60mm]{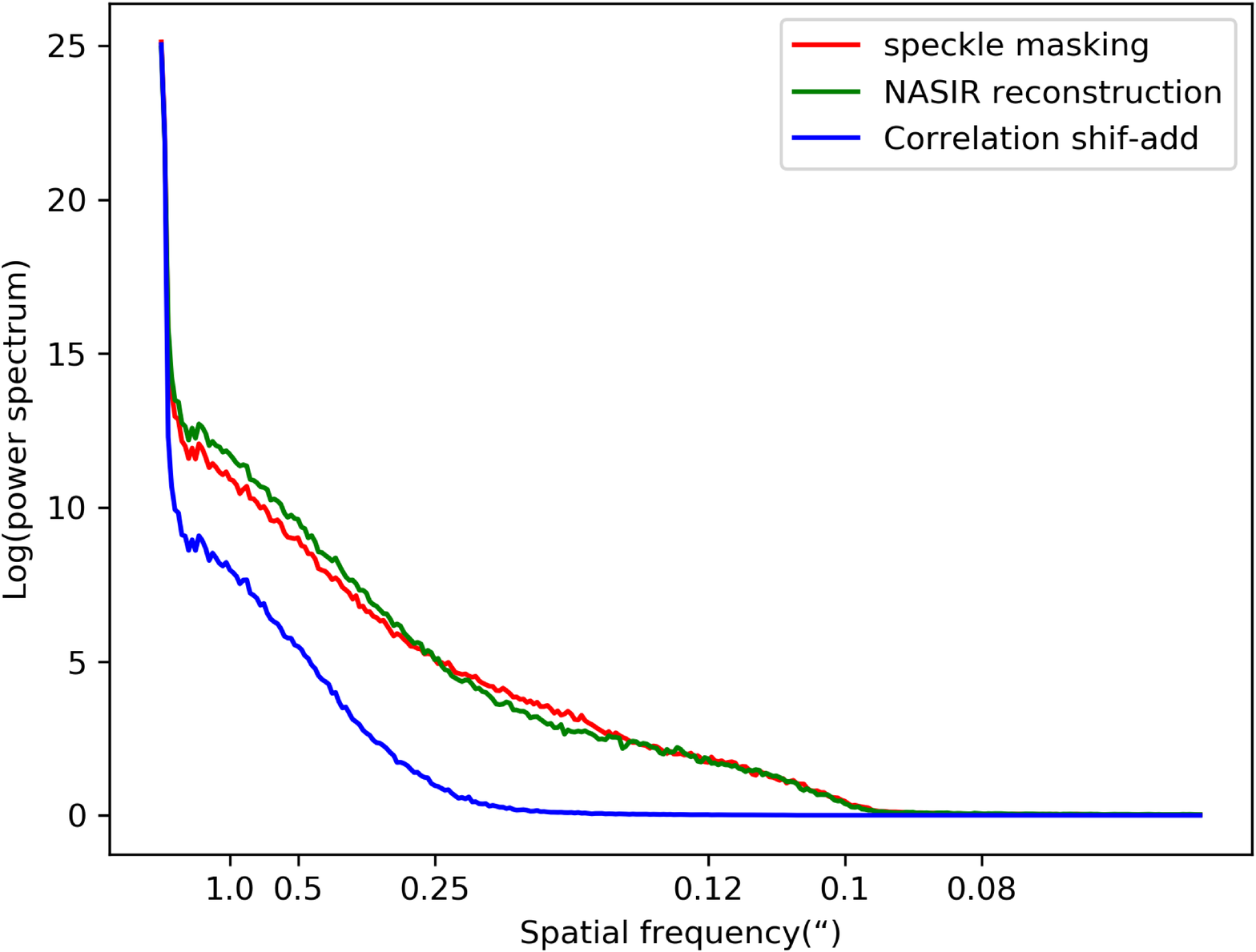}
  \end{minipage}%
  \begin{minipage}[t]{0.495\textwidth}
  \centering
   \includegraphics[width=60mm]{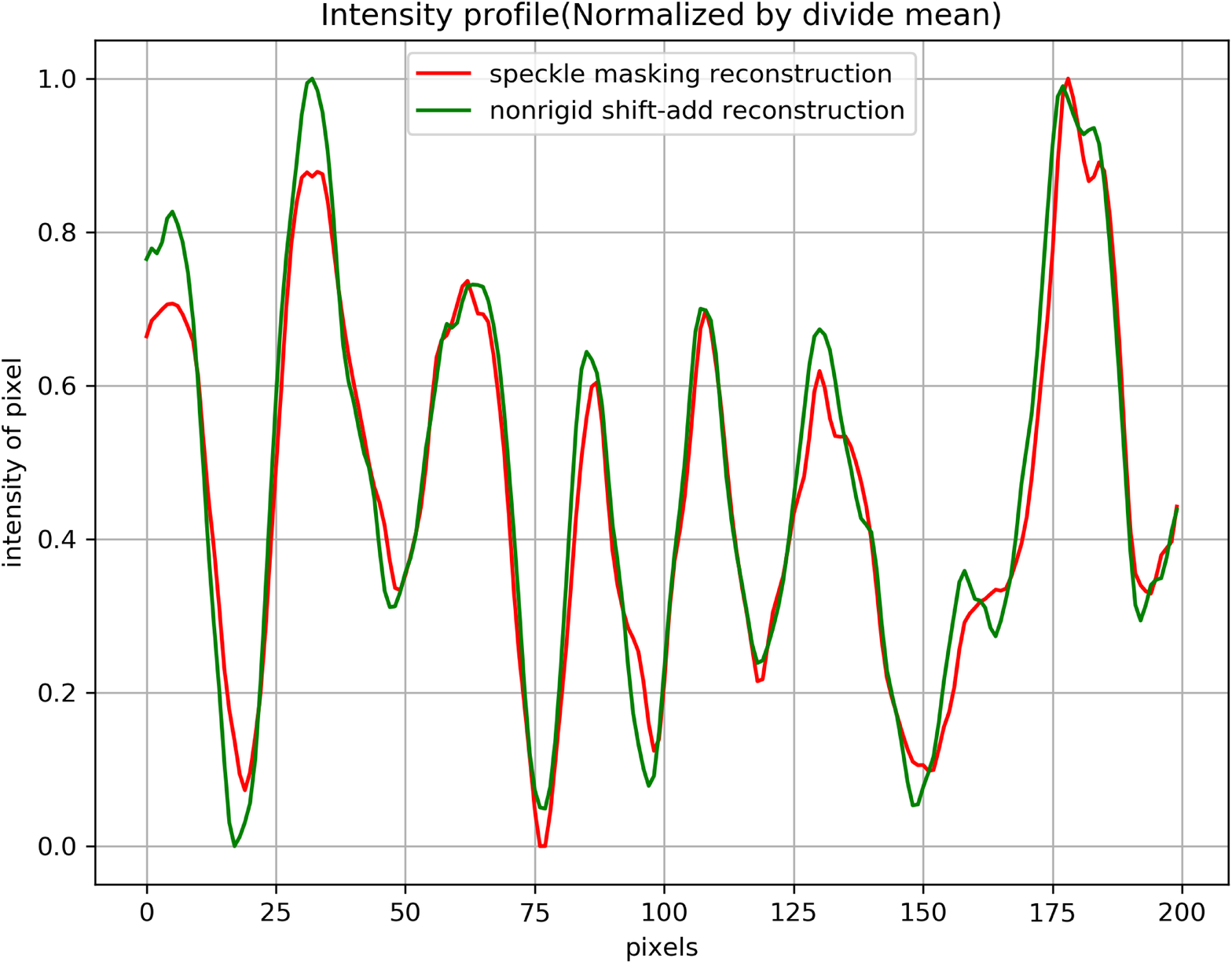}
  \end{minipage}%
  	  \caption{\label{Fig2}{Power spectrum curves (left) and intensity profile(right, the red line region in Fig. 1 (a))) of images in Fig1.}}
\end{figure}

\par To further reveal the performance of NASIR, we use the following quantitative metrics to compare and analyze the reconstruction results of \textit{speckle masking} and NASIR:
\par \textbf{(1) Image correlation coefficient} (\emph{r}): The correlation coefficient is used to compare the degree of linear correlation between reconstructed images, which is defined as:
\begin{equation}\label{cc}
\emph{r}(\boldsymbol{X},\boldsymbol{Y})=\frac{Cov(\boldsymbol{X},\boldsymbol{Y})}{\sqrt{Var[\boldsymbol{X}]\cdot{Var[\boldsymbol{Y}]}}}
\end{equation}
Here, $Cov(\boldsymbol{X},\boldsymbol{Y})$ is the covariance of the pixel intensity of the two images $\boldsymbol{X}$ and $\boldsymbol{Y}$, $Var[\boldsymbol{X}]$ and $Var[\boldsymbol{Y}]$ are the variances of $\boldsymbol{X}$ and $\boldsymbol{Y}$ respectively.
\par \textbf{(2) Structural SIMilarity} (\emph{SSIM}) \citep{wang2004image}: \emph{SSIM} is an index that measures the similarity of two images. Given two images $\boldsymbol{X}$ and $\boldsymbol{Y}$, their structural similarity can be defined in the following way:
\begin{equation}\label{ssim}
\emph{SSIM}(\boldsymbol{X},\boldsymbol{Y})=\frac{(2\mu_x\mu_y+c_1)\cdot{(2\sigma_{xy}+c_2)}}{(\mu_x^2+\mu_y^2+c_1)\cdot{(\sigma_x^2+\sigma_y^2+c_2)}}
\end{equation}
Where $\mu_x$ is the mean of $\boldsymbol{X}$, $\mu_y$ is the mean of $\boldsymbol{Y}$, $\sigma_x$ is the variance of $\boldsymbol{X}$, $\sigma_y$ is the variance of $\boldsymbol{Y}$, and $\sigma_{xy}$ is the covariance between $\boldsymbol{X}$ and $\boldsymbol{Y}$. $c_1=(k_1L)$, $c_2=(k_2L)$ are constants used to maintain stability. $L$ is the range level of pixel values. $k_1=0.03$, $k_2=0.03$. The range of structural similarity is -1 to 1. When the two images are the same, the value of \emph{SSIM} is equal to 1.
\par The structural similarity index defines structural information from the perspective of image composition as being independent of brightness and contrast, reflecting the properties of the object structure in the scene, and modeling distortion as a combination of three different factors, brightness, contrast, and structure. The mean is used as an estimate of brightness, the standard deviation is used as an estimate of contrast, and covariance is used as a measure of structural similarity.
\par \textbf{(3) Coefficient of Variation of Intensity Profile} (\emph{CVoIP}): We use the coefficient of variation of the intensity profile of the reconstructed image, which is the ratio of the standard deviation of the intensity profile to its mean, to measure the difference in image contrast.
\par A quantitative comparison of \textit{speckle masking} with NASIR is shown in Table 2, indicating that the reconstructed images of the two methods have high correlation and structural similarity.
\begin{table}
\begin{center}
\caption{Quantitative comparison of the reconstruction results of the four groups of test data (Table 1).}
\label{T2-simple}
\begin{tabular}{ccccc}     
  \hline                   
Data set No. & \emph{r} & \emph{SSIM} & \qquad \qquad \emph{CVoIP}  \\
     &      &     & ~\emph{speckle masking} & NASIR \\
  \hline
1 & 0.9802 & 0.9561  & 0.5022 & 0.5359 \\
2 & 0.9489 & 0.8887  & 0.3580 & 0.4078 \\
3 & 0.9129 & 0.8154  & 0.3636 & 0.4911 \\
4 & 0.5182 & 0.5417  & 0.4608 & 0.4795 \\

  \hline
\end{tabular}
\end{center}
\end{table}
Therefore, the visual quality (Fig.1), the consistency of the power spectrum curve (Fig.2), and the quantitative comparison index of Table 2 (first row) show that when seeing is good, NASIR can achieve reconstruction quality close to the speckle masking method.
\par The reconstruction results of the H$\alpha$-band data further verify the effectiveness of the NASIR for high-resolution reconstruction of the speckle images when the seeing is better. Fig.3 shows the reconstructed images of the three algorithms for data set 2 in Table 1. The quantitative comparison result is in the second row in Table 2. The power spectrum curves and the intensity profile curves are shown in Fig.4.

\begin{figure}[ht]
  \begin{minipage}[t]{0.495\linewidth}
  \centering
   \includegraphics[width=60mm]{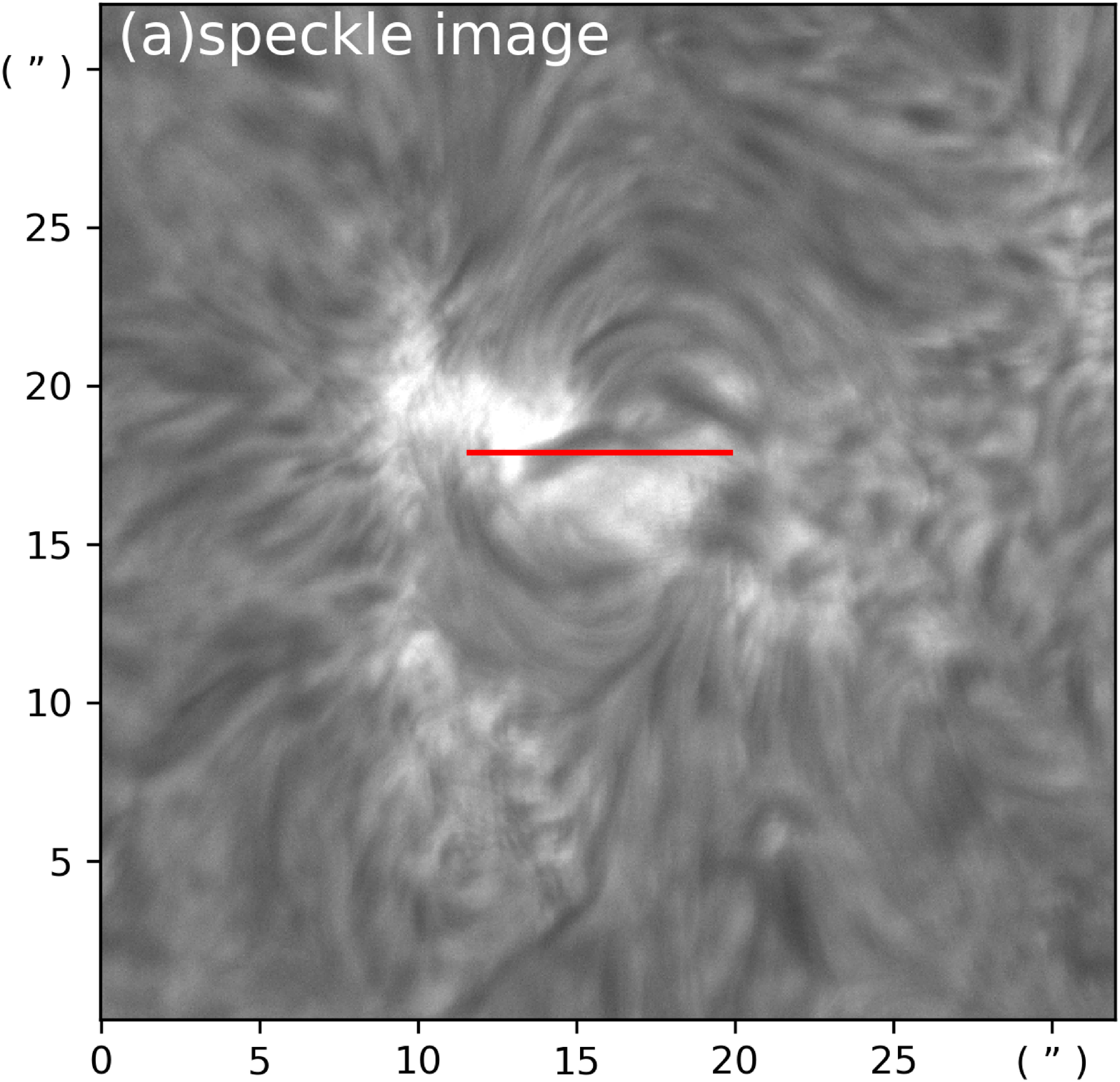}
  \end{minipage}%
  \begin{minipage}[t]{0.495\textwidth}
  \centering
   \includegraphics[width=60mm]{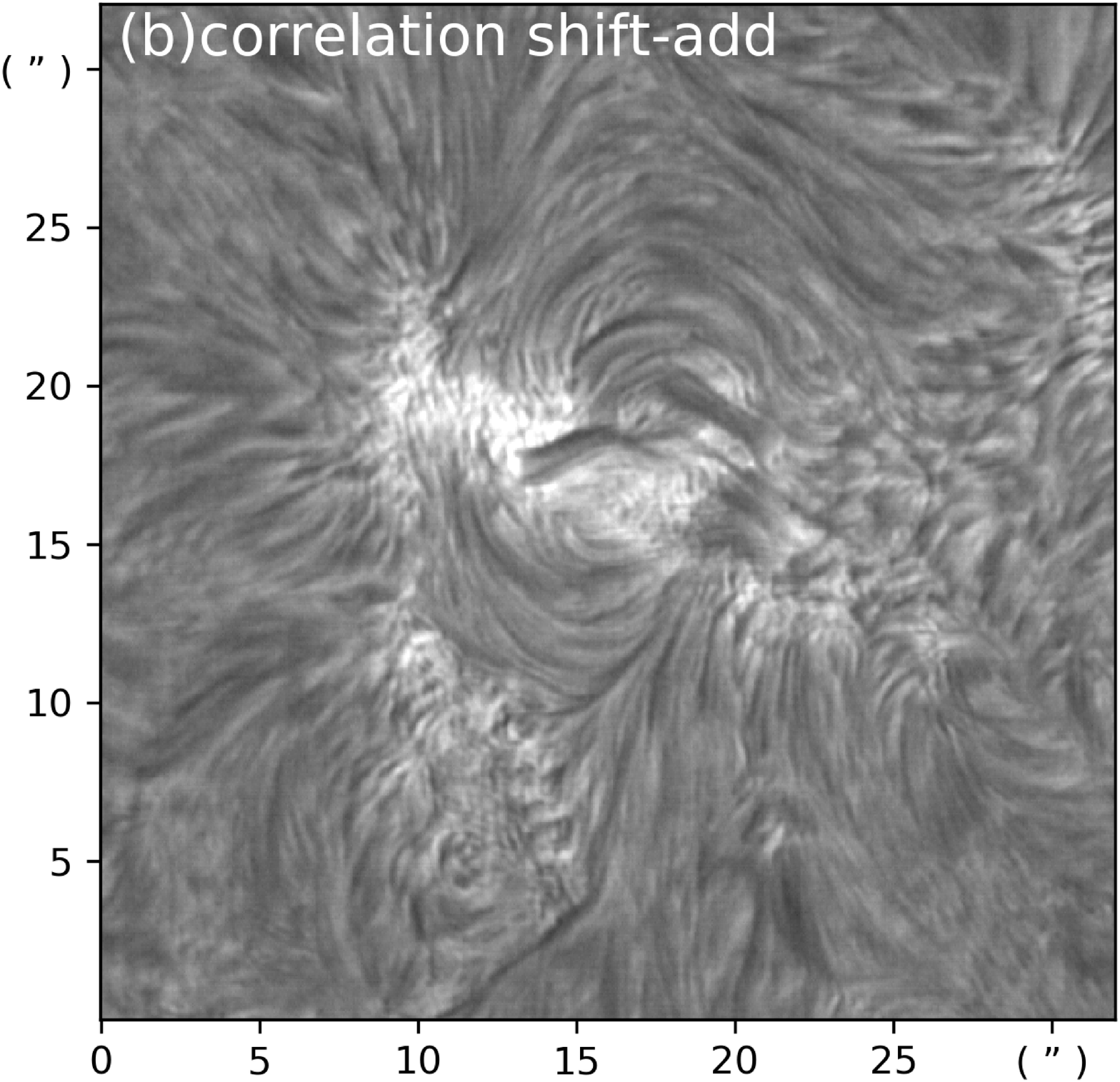}
  \end{minipage}%

  \begin{minipage}[t]{0.495\textwidth}
  \centering
     \includegraphics[width=60mm]{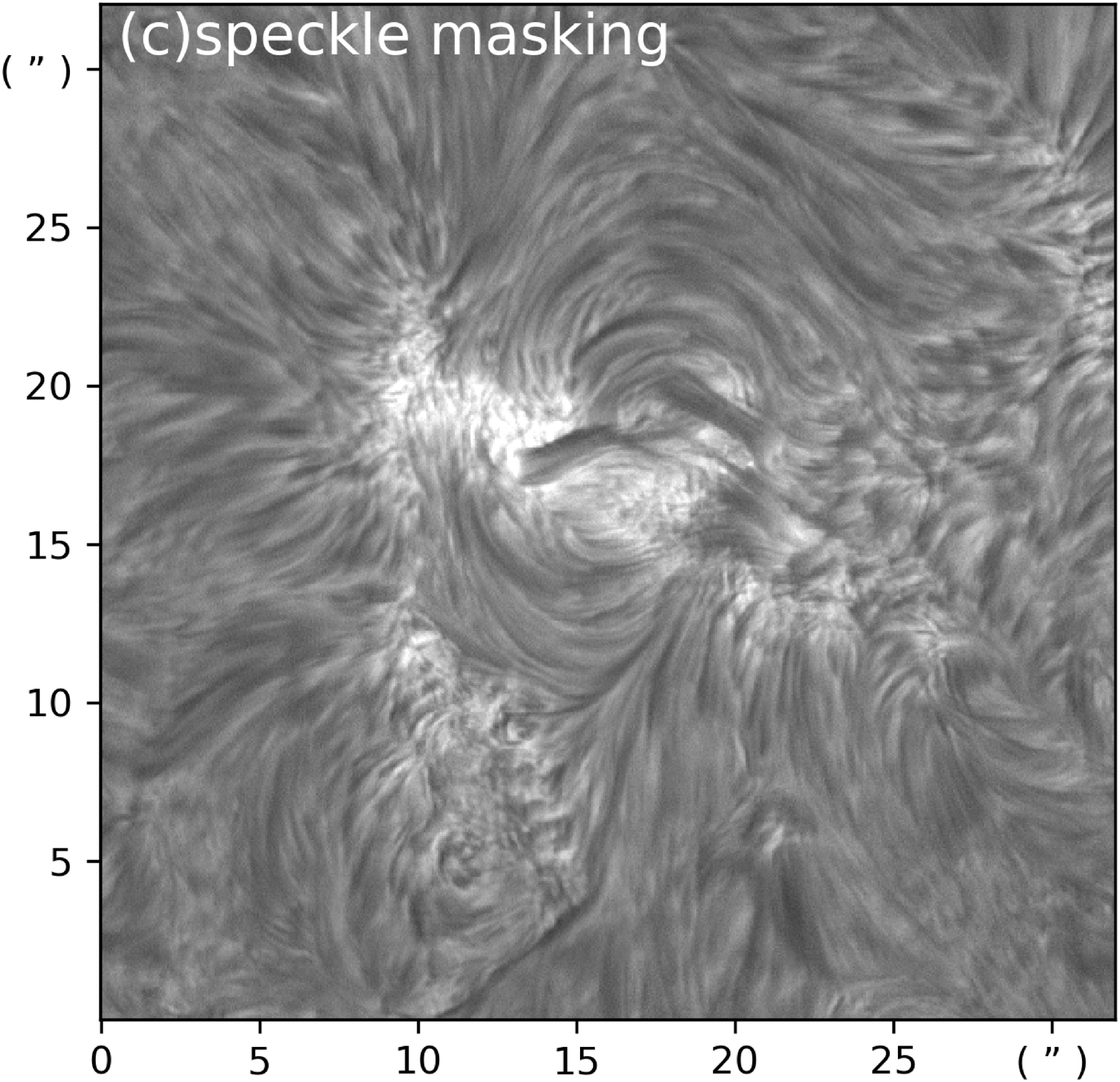}
  \end{minipage}%
  \begin{minipage}[t]{0.495\textwidth}
  \centering
   \includegraphics[width=60mm]{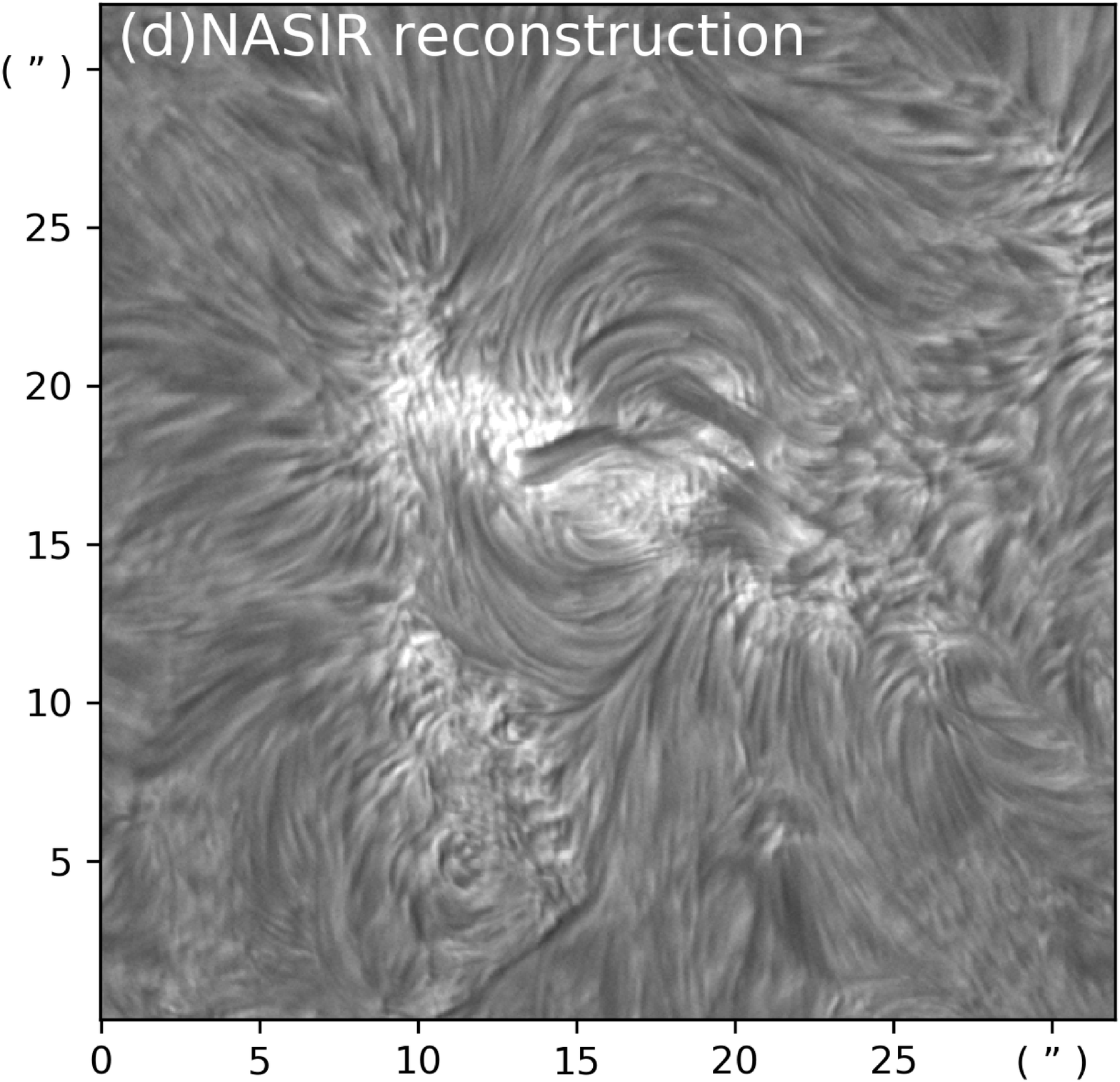}
  \end{minipage}%
  \caption{\label{Fig3} Reconstruction results and power spectrum of NVST H$\alpha$ 2019-10-04T02:59:05 (active area 12749).}
\end{figure}

\begin{figure}[ht]
  \begin{minipage}[t]{0.495\linewidth}
  \centering
   \includegraphics[width=60mm]{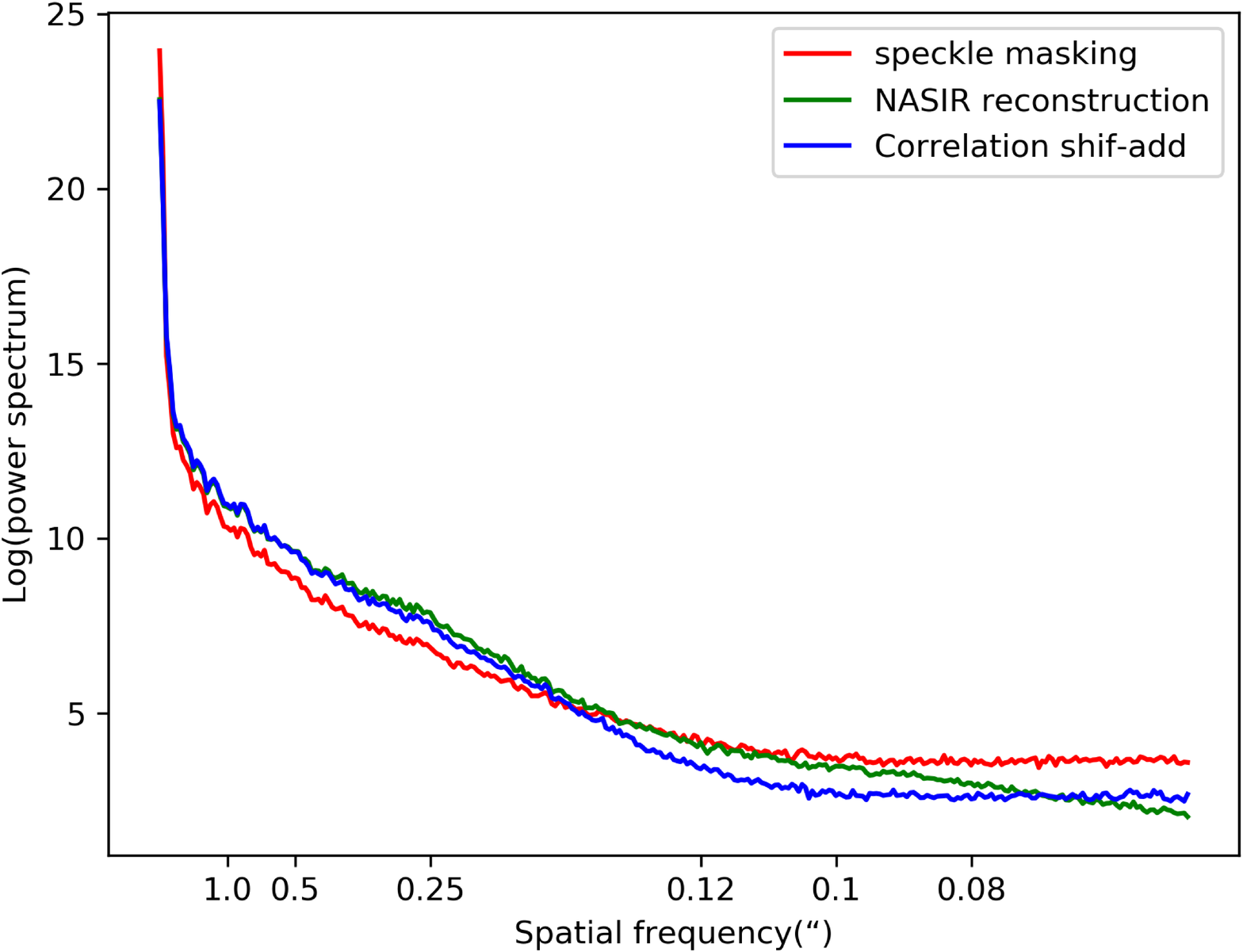}
  \end{minipage}%
  \begin{minipage}[t]{0.495\textwidth}
  \centering
   \includegraphics[width=60mm]{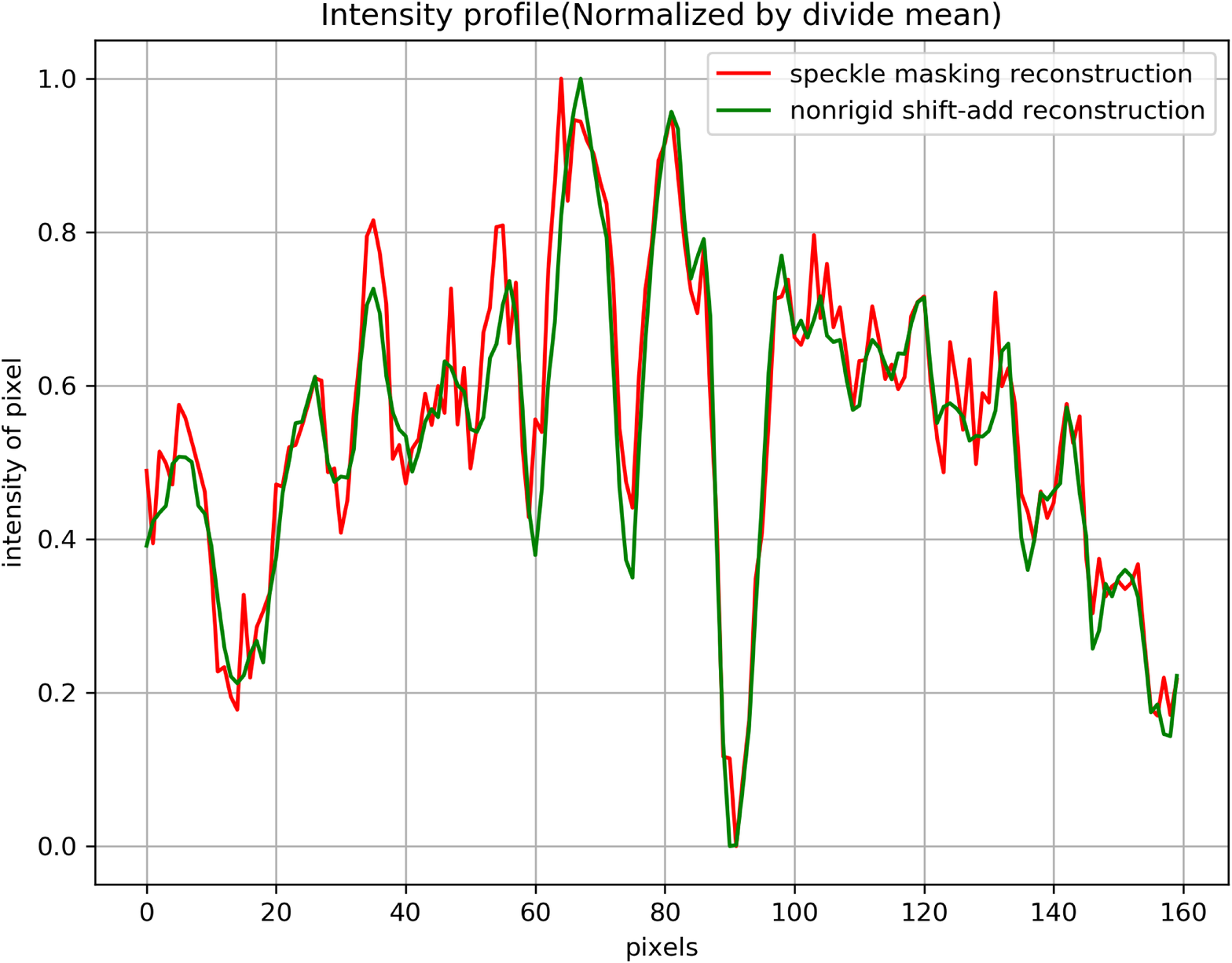}
  \end{minipage}%
  	  \caption{\label{Fig4}{Power spectrum curves (left) and intensity profile(right) of images in Fig3. }}
\end{figure}

\subsection{Comparison of Reconstruction Performance with Slightly Poor Seeing} 
  \label{S-Comparison of Reconstruction Performance with Slightly Poor Seeing}

  Fig.5 shows the reconstruction results of dataset 3 ($\boldsymbol{r_0}$=5.88cm). It can be seen that as the seeing decreases, the recursive cumulative error of the phase of \textit{speckle masking} has an aggravated influence on the reconstruction, resulting in stray structures in the reconstructed image, while NASIR can better maintain the characteristics of granules. The reconstruction results of the two algorithms still have high correlation and structural similarity, but the CVoIP values reflect the higher contrast of the reconstructed images from NASIR. The third row of Table 2 gives the comparative quantitative indicators.

\begin{figure}[ht]
  \begin{minipage}[t]{0.495\linewidth}
  \centering
   \includegraphics[width=60mm]{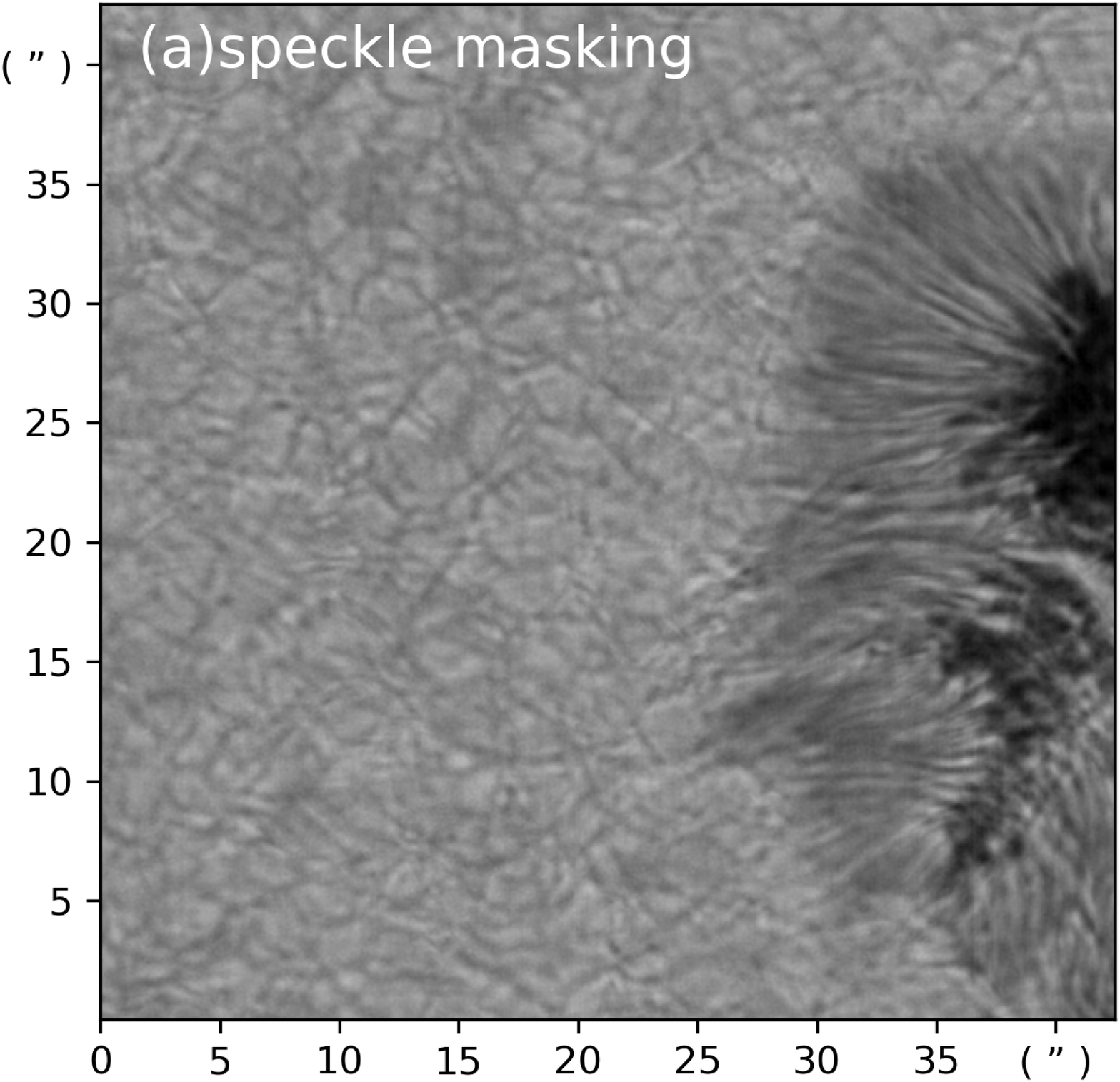}
  \end{minipage}%
  \begin{minipage}[t]{0.495\textwidth}
  \centering
   \includegraphics[width=60mm]{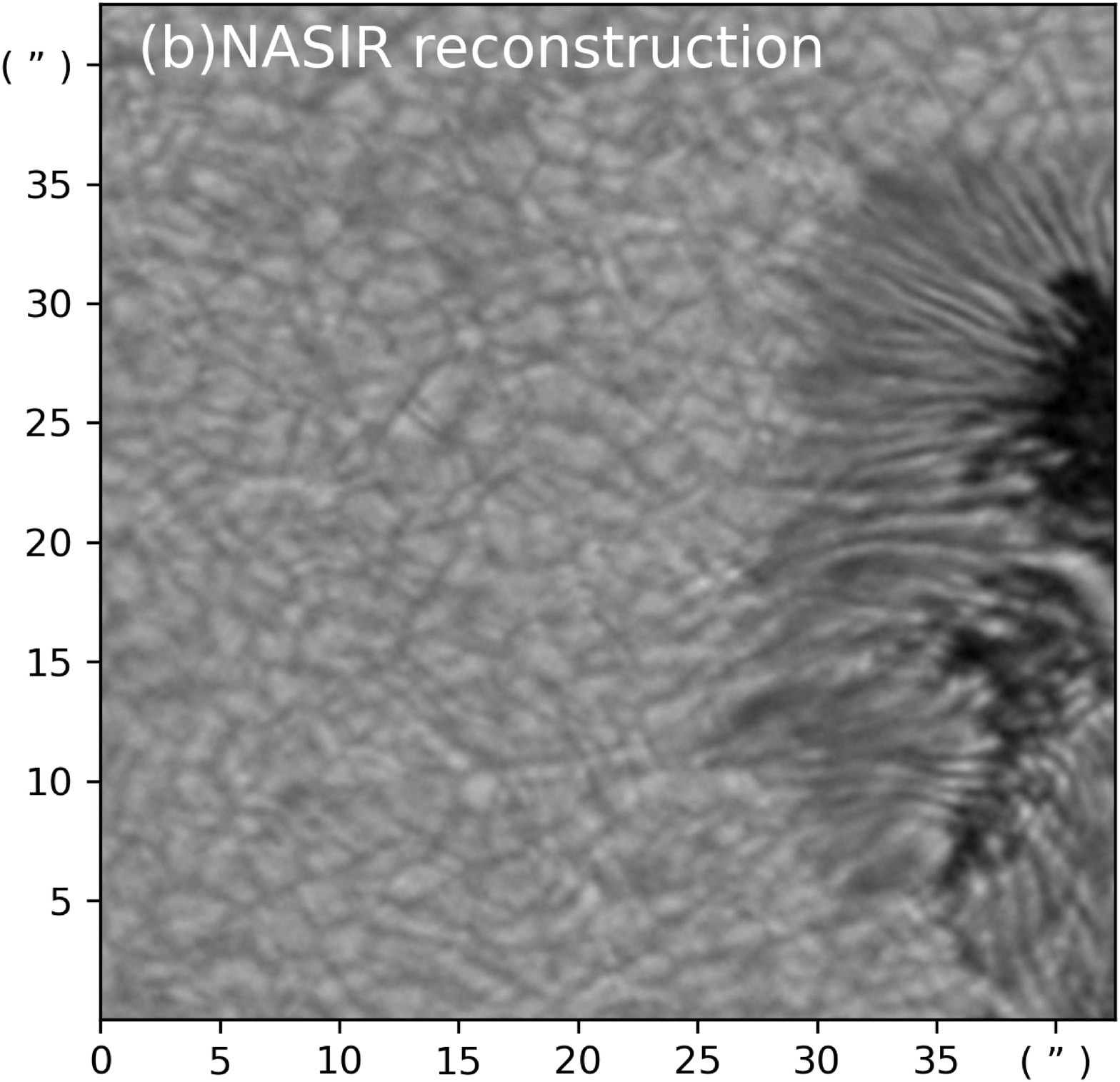}
  \end{minipage}%
  	  \caption{\label{Fig5}{Comparison of reconstruction results when $\boldsymbol{r_0}$=5.88cm, (a) \textit{speckle masking} (b) NASIR.}}
\end{figure}

Fig.6 shows the reconstruction results  of data set 4 ($\boldsymbol{r_0}$=4.09cm) in Table 1. As we can see, with the further decrease in seeing, the correlation coefficient \emph{r} and \emph{SSIM} of the images reconstructed by the two algorithms start to decrease significantly (the fourth row of Table 2), indicating a clear difference in the reconstructed images. The phase recurrence error of \textit{speckle masking} seriously affects the reconstruction quality, the reconstructed image has been destroyed by the pseudo-structure, while NASIR can still obtain low-frequency structures to a certain degree.

\begin{figure}[ht]
  \begin{minipage}[t]{0.495\linewidth}
  \centering
   \includegraphics[width=60mm]{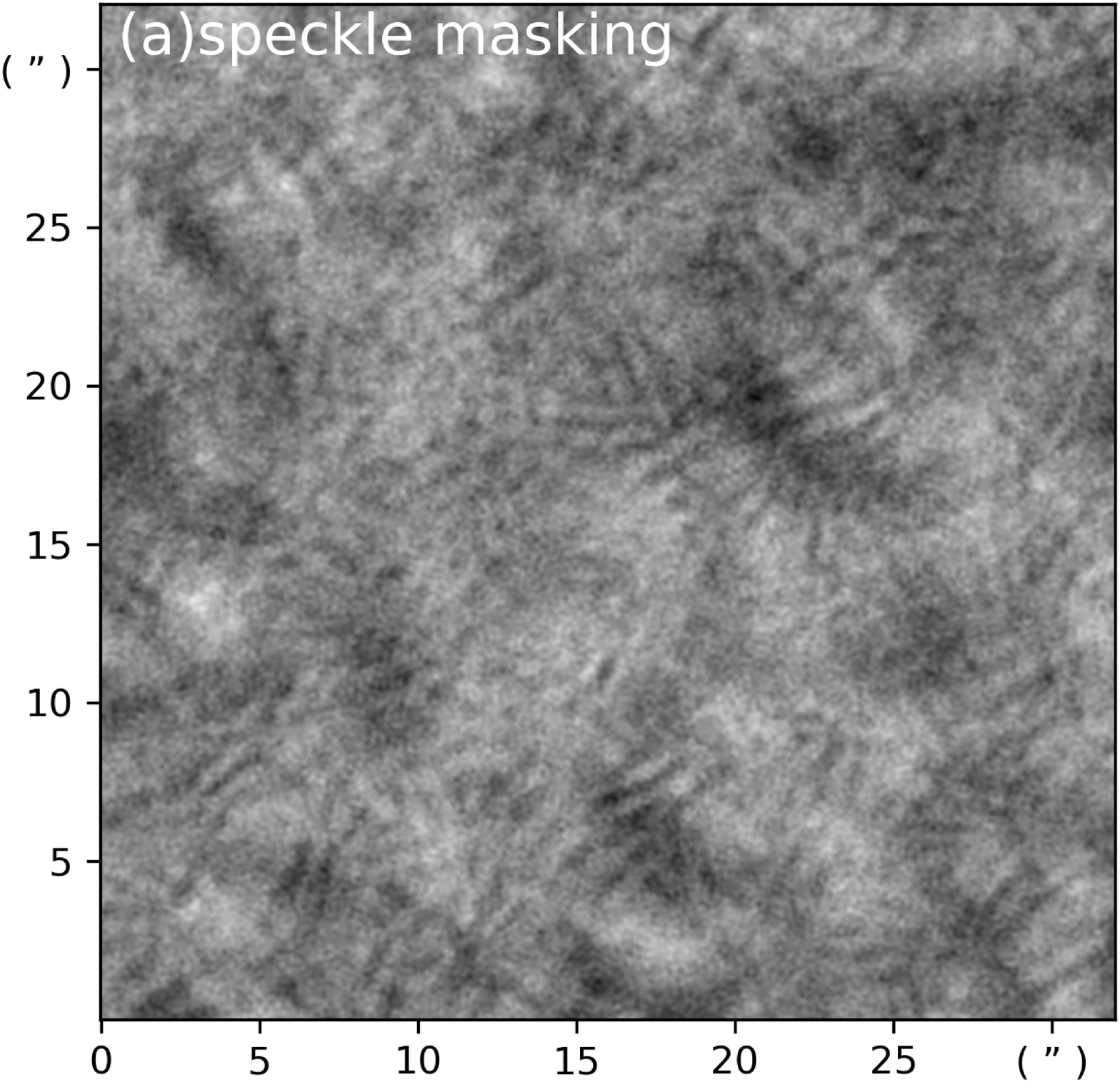}
  \end{minipage}%
  \begin{minipage}[t]{0.495\textwidth}
  \centering
   \includegraphics[width=60mm]{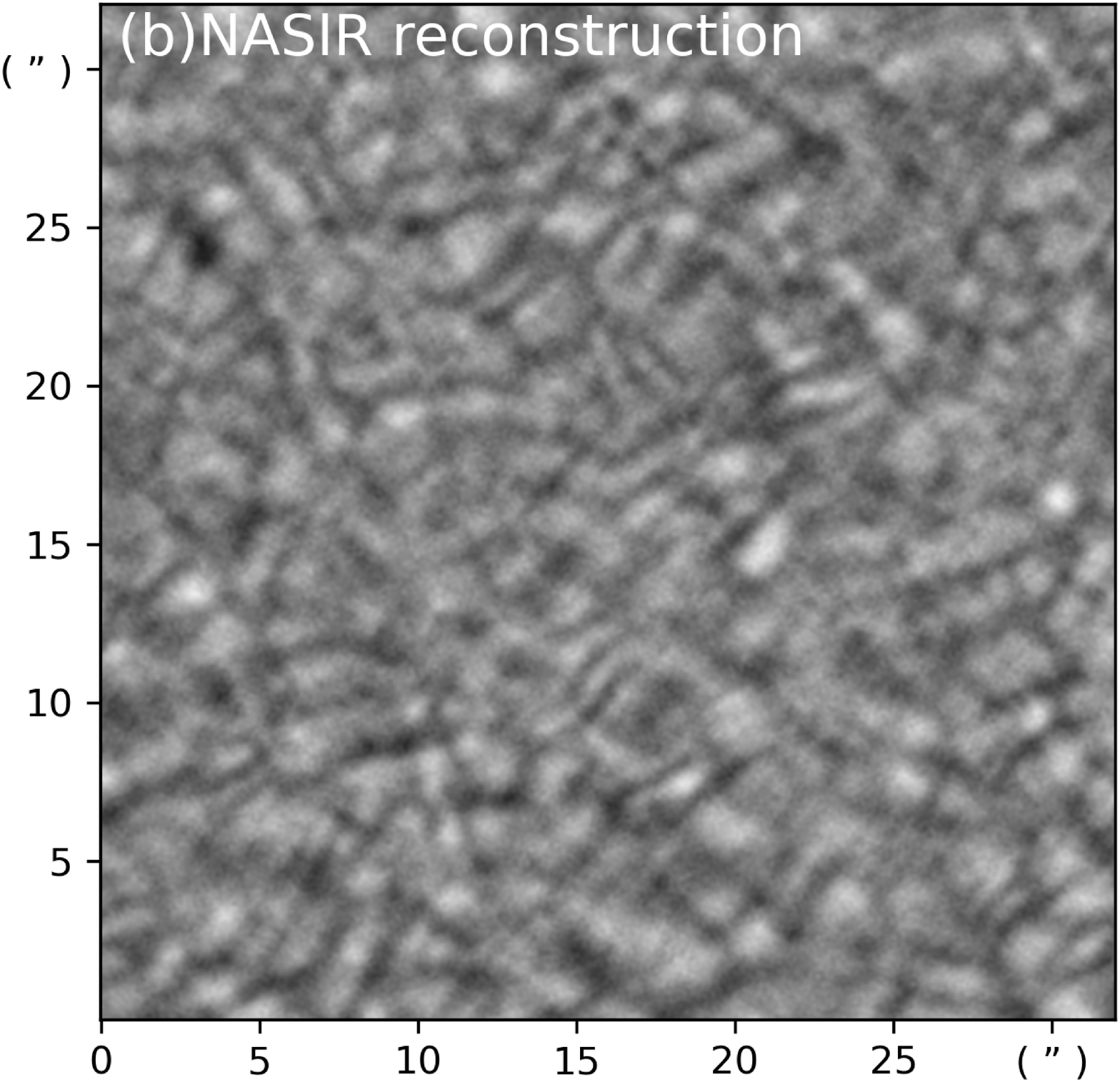}
  \end{minipage}%
  	  \caption{\label{Fig6}{Comparison of reconstruction results when $\boldsymbol{r_0}$=4.09cm, (a) \textit{speckle masking}, (b) NASIR.}}
\end{figure}

\subsection{Computational Time-consuming} 
  \label{S-Computational Time-consuming}
NASIR is a spatially parallel algorithm that avoids the multi-path phase recursion, block reconstruction, and stitching, so its average computation time is about 48.7$\%$ of \textit{speckle masking}.

\section{Discussion}
     \label{S-Discussion}

The core idea of NASIR is to recover the phase distortion of different frequency components through non-rigid (non-linear) geometric correction. It is a fast and robust method for high-resolution statistical reconstruction.

\subsection{Features of NASIR} 
  \label{S-Advantages of NASIR}

\par NASIR performs phase correction through pixel-by-pixel non-rigid alignment, which is the affine correspondence between pixel feature descriptors.
Feature descriptors are polynomial fit coefficients of the intensities of pixels and their neighborhoods. So non-rigid alignments are not limited by spatial invariance (isoplanatic angle). In addition, the continuous update of the reference image \textbf{RI} by NASIR can make the low-frequency structure tend to be consistent during the iterative alignment process, so mitigating the geometric distortions between isoplanatic angles. Therefore, NASIR can directly reconstruct the field of view at a time without block-wise reconstruction.

\par NASIR avoids the computation of high-order statistics and complex multi-path phase recurrence, and the computation time is only about half that of \textit{speckle masking}.

\par NASIR estimates the displacement field for each pixel of the field of view in parallel, which can be immune to recursive error accumulation.
\par Band-pass filtering of the raw speckle images suppresses noise interference beyond the diffraction limit, allowing NASIR to reconstruct the modulus and phase of the mid-frequency structure more efficiently.

\subsection{Further Improvements to NASIR} 
  \label{S-Further Improvements to NASIR}
While achieving successful experiments, NASIR deserves further optimization and improvement in the following areas:
\par (1) Optimization of displacement field-intensity distribution model. Considering the convenience of analysis and calculation, NASIR adopts polynomial expansion to approximate the intensity distribution of the image area. It may differ from physical models of actual intensity changes caused by atmospheric turbulence. Therefore, establishing a speckle image intensity distribution-displacement representation model more in line with physics is one of the issues worthy of further study.
\par (2) Adaptive selection of neighborhood size. NASIR uses the intensity distribution model of the neighborhood as a feature descriptor and estimates the displacement through the matching between feature descriptors. A larger neighborhood size can improve the matching stability, but it will produce more smoothing effects and reduce the reconstruction quality of high-frequency structures. Conversely, a smaller neighborhood size can preserve high-frequency details. But in the case of poor vision and high noise, it is easy to cause mismatch and reduce the accuracy of displacement estimation. Therefore, a more appropriate size should be chosen according to the atmospheric coherence length for better alignment performance.
\par (3) Design and implementation of seeing-based band-pass filters. NASIR uses band-pass filtered speckle images for alignment, essentially phase correction based on the alignment of object structures within the diffraction limit. Therefore, seeing-based adaptive band-pass filter design is also one of the fundamental issues to consider.
\par (4) Improve alignment and update iteration strategies. Non-rigid alignment based on speckle imaging requires a reference image. Iterative improvement of reference image quality is crucial for phase reconstruction. Empirically, the number of iterations for NASIR to update the reference image is between 3$\sim$5 times, therefore, it is necessary to make more reasonable choices and judgments on iteration and convergence based on seeing $\boldsymbol{r_0}$.

\section{Conclusion}
 \label{S-Conclusion}
NASIR is a novel high-resolution reconstruction method for speckle imaging that combines classical statistical reconstruction methods and computer vision techniques. Reconstruction results from ground-based telescope observations demonstrate the potential of NASIR in several ways:
\par \textbf{Flexibility}. NASIR can directly reconstruct the full field of view without sub-block reconstruction and stitching, avoiding stitching errors and discontinuities between stitched sub-blocks.
\par \textbf{Robustness}. When seeing is good, the reconstruction quality of NASIR is close to speckle masking. When seeing decreases,  NASIR is more stable.
\par \textbf{Fastness}. NASIR does not perform phase recursion and avoids block reconstruction/stitching operations. Computation time is reduced by more than 50$\%$ compared to \textit{speckle masking}.
\par In conclusion, although NASIR does not have theoretical accuracy guarantees like \textit{speckle masking}. However, experimental results show that NASIR is still a convenient, robust, and fast method for high-resolution solar image reconstruction. We believe that it will play an important role in many applications such as quick look and data filtering.

\begin{acknowledgements}
We would like to thank the NVST team for high cadence data support.
This work is sponsored by the National Natural Science Foundation of China (NSFC) under
the grant numbers (11873027, U2031140, 12073077, 11833010, 11973088), and West Light
Foundation of the Chinese Academy of Sciences (Y9XB01A, Y9XB019).
\end{acknowledgements}

\label{lastpage}

\bibliographystyle{raa}
\bibliography{ms2022-0154}

\end{document}